\documentclass[epj-spec]{svjour}
\usepackage{amsmath}
\usepackage{amssymb}
\usepackage{bm}
\renewcommand{\vec}[1]{\boldsymbol{#1}}
\newcommand{\tensor}[1]{{\boldsymbol{\mathit {#1}}}}
\newcommand{\Nabla}{\vec{\nabla}}
\newcommand{\Lnabla}{\overset{\leftarrow}{\Nabla}}

\newcommand{\ket}[1]{\left| #1 \right\rangle}
\newcommand{\EW}[1]{\langle #1 \rangle}

\newcommand{\D}{\mathrm{d}}
\renewcommand{\Im}{{\rm Im}\,}
\renewcommand{\Re}{{\rm Re}\,}

\newcommand{\fo}[1]{\underline{\hat{\mathbf{#1}}}}
\newcommand{\fon}[1]{\underline{\mathbf{#1}}}
\newcommand{\ot}{}         
\usepackage{graphics}
\begin{document}
\title{QED in arbitrary linear media: amplifying media}
\author{Christian Raabe%
\thanks{\email{C.Raabe@tpi.uni-jena.de}}
\and Dirk-Gunnar Welsch%
}
\institute{Theoretisch-Physikalisches Institut,
Friedrich-Schiller-Universit\"at Jena, Max-Wien-Platz 1,
D-07743 Jena, Germany}
\abstract{%
Recently, we have developed
a unified approach to QED in
arbitrary linearly responding media
in equilibrium---media that give
rise to absorption
[Phys. Rev. A \textbf{75}, (2007) 053813].
In the present paper we show that, under appropriate
conditions, the theory can be
quite naturally
generalized to amplifying media
the effect of which is
described within the framework of linear response theory.
We discuss the limits of validity of the generalized
theory and make contact with earlier quantization schemes
suggested for the case of linearly and locally responding
amplifying dielectric-type media.
To illustrate the theory, we present the 
electromagnetic-field correlation functions that
determine the Casimir force in the presence of
amplifying media.%
}
\maketitle
\section{Introduction}
\label{sec2-0}

In macroscopic electrodynamics, the 
effect of an arbitrary linearly responding
medium in equilibrium can be described by means 
of a complex conductivity tensor
of the unperturbed medium left to its own resources.
It is well known that any
electromagnetic excitation
in such a medium
eventually dissipates due to unavoidable
absorption
in the whole frequency range.
On the other hand, if a medium has been brought out of
equilibrium, it may act, in a certain frequency range, as an amplifier
until after some relaxation time
it will have reached the equilibrium
state again.
A detailed description
of the temporal evolution
of the medium properties
during such transient processes is a highly
difficult problem in general.
However, when---%
by implementation of a suitable energy supply mechanism---%
a medium is pumped in such a way that
an (externally controlled) quasi-stationary
regime is established and maintained for a sufficiently long time,
one may describe the effect of the amplifying
medium by assigning to it
a
conductivity 
tensor.
The properties
of such an amplifying medium
are thereby regarded
as being fixed---an assumption
which is of course only
justified
%
in a time interval within which
significant changes of these properties
can be ignored (for the purpose at hand).
Clearly, the
conductivity tensor that describes 
the linear response of an amplifying medium
can be expected to have all the
properties known from an absorbing medium,
except that it is no longer
the kernel of a positive-definite integral
operator.

Both absorption and amplification introduce
additional noise into the system---noise that must be carefully
taken into account in quantum electrodynamics (QED).
Recently, we have developed a 
unified approach to macroscopic QED in arbitrary linearly
responding media
described by a conductivity tensor for absorbing
media \cite{RaabeC052007}. In the present paper, we extend this 
quantization scheme to allow also for amplifying media.
In Sec.~\ref{sec2-1}, we
outline the main aspects of the scheme
and indicate at which points
modifications are necessary when a medium becomes
amplifying.
The extension of the theory
to amplifying media is then given
in Sec.~\ref{sec1-5}, where we also make
contact with earlier quantization schemes for the electromagnetic field
in (locally responding dielectric) amplifying media
\cite{ScheelS071998,MatloobR031997}.
In this context, we also address the problem of
the necessity of a modification of the Hamiltonian,
which has not been considered in
Refs.~\cite{ScheelS071998,MatloobR031997}.
It is worth noting that without
this modification the
field operators, whose positive-frequency components
are associated with both annihilation and creation 
operators in
the case of amplifying media, would fail to satisfy Maxwell's equations.
To illustrate the theory, 
and to make contact with recent work on the Casimir force
in the presence of linearly amplifying media~\cite{LeonhardtU082007}, 
we present the vacuum-field correlation 
functions that are relevant to the Casimir force.
The example shows that, as expected, the form
of the field correlation functions changes significantly, 
compared to the familiar form observed in the case 
of equilibrium media, when 
amplifying media are involved---a point that
has not been taken into account in Ref.~\cite{LeonhardtU082007}. 
In Sec.~\ref{sec1-6}, we discuss in detail the limits
of validity of the theory.
Finally, we give a summary and some concluding remarks
in Sec.~\ref{sec1-7}.
%


\section{Sketch of the quantization scheme}
\label{sec2-1}
It is well-known that Maxwell's equations
imply that the electric field in the frequency domain,
$\fon{E}(\mathbf{r},\omega)$,
obeys, in the presence of matter, the (classical) equation
\begin{equation}
\label{Max5}
\Nabla\times\Nabla\times\fon{E}(\mathbf{r},\omega)
-\frac{\omega^2}{c^2}\,\fon{E}(\mathbf{r},\omega)
=i\mu_{0}\omega \fon{j}(\mathbf{r},\omega),
\end{equation}
where
$\fon{j}(\mathbf{r},\omega)$
is the total
current density,
%
i.\,e., the current density that covers
all the matter (on a chosen length scale).
If $\fon{j}(\mathbf{r},\omega)$ 
can be attributed entirely to a medium 
whose internal atomistic structure need not be resolved, 
we may assume, within the framework of linear response theory,
the constitutive relation
\begin{equation}
\label{eq8-1}
\fon{j}(\mathbf{r},\omega)=\int
\D^3r'\,\tensor{Q}(\mathbf{r},\mathbf{r}',\omega)
\cdot\fon{E}(\mathbf{r'},\omega)
+\fon{j}_{\mathrm N}(\mathbf{r},\omega),
\end{equation}
where $\tensor{Q}(\mathbf{r},\mathbf{r}',\omega)$
is the complex, macroscopic
conductivity tensor
of the unperturbed medium
in the frequency domain
\cite{KuboNonEqStat,MelroseBook},
and $\fon{j}_{\mathrm N}(\mathbf{r},\omega)$ is
a Langevin noise source.
Equation (\ref{eq8-1}) covers all the
possible features of a linear medium
(in particular, temporal as well as spatial dispersion).
According to Onsager's
reciprocity theorem
\cite{KuboNonEqStat,MelroseBook},
the conductivity tensor
fulfills the reciprocity relation
$Q_{ij}(\mathbf{r},\mathbf{r}',\omega)$
$\!=Q_{ji}(\mathbf{r}',\mathbf{r},\omega)$, 
which we will adopt throughout the paper.
For chosen $\omega$, $\tensor{Q}(\mathbf{r},\mathbf{r}',\omega)$ can be assumed
to be the integral kernel of a reasonably well-behaved integral
operator acting on vector functions in
position space.
In particular,
in the spatially non-dispersive limit, it
may become a
(qua\-\mbox{si-)l}o\-cal integral kernel,
i.\,e.,
a linear combination of $\delta$-functions and
their derivatives.
It should be noted that the decomposition
of the conductivity tensor into real and imaginary parts,
$\tensor{Q}(\mathbf{r},\mathbf{r}',\omega)$ $\!=$
$\!\Re\tensor{Q}(\mathbf{r},\mathbf{r}',\omega)$ $\!+$
$\!i\,\Im\tensor{Q}(\mathbf{r},\mathbf{r}',\omega)$,
corresponds, due to the reciprocity of
$\tensor{Q}(\mathbf{r},\mathbf{r}',\omega)$, to
the decomposition of the associated operator
into Hermitian and anti-Hermitian parts,
\begin{align}
\label{eq2}
\tensor{\sigma}(\mathbf{r},\mathbf{r}',\omega)
\equiv \Re\tensor{Q}(\mathbf{r},\mathbf{r}',\omega)
={\textstyle\frac{1}{2}}
\bigl[\tensor{Q}(\mathbf{r},\mathbf{r}',\omega)+
\tensor{Q}{^{+}}(\mathbf{r}',\mathbf{r},\omega)\bigr],
\end{align}
\begin{align}
\label{eq3}
\tensor{\tau}(\mathbf{r},\mathbf{r}',\omega)
\equiv
\Im\tensor{Q}(\mathbf{r},\mathbf{r}',\omega)
={\textstyle\frac{1}{2i}}
\bigl[\tensor{Q}(\mathbf{r},\mathbf{r}',\omega)
-\tensor{Q}{^{+}}(\mathbf{r}',\mathbf{r},\omega)\bigr].
\end{align}
Throughout the paper, the
superscripts~${}^\mathsf{T}$ and~${}^{+}$ are used
to indicate transposition and Hermitian conjugation
with respect to tensor indices.
Since $\tensor{Q}(\mathbf{r},\mathbf{r}',\omega)$
is the temporal Fourier transform of a response
function 
\cite{KuboNonEqStat,NussenzweigBook,LanLifStat},
%
it is analytic in the upper complex $\omega$ half-plane,
fulfills the Kramers--Kronig (Hilbert transform) relations,
and satisfies the Schwarz reflection principle
$\tensor{Q}{^{\ast}}(\mathbf{r,r'},\omega)$
$\!=\tensor{Q}(\mathbf{r,r'},-\omega^{\ast})$.
Since an unperturbed medium left to its own resources 
is necessarily an absorbing one,
the operator associated with the integral kernel
$\tensor{\sigma}(\mathbf{r},\mathbf{r}',\omega)$ is
%
%
a positive definite operator
(see, e.\,g., Refs.~\cite{KuboNonEqStat,MelroseBook}),
i.\,e., for any (quadratically integrable) vector
function $\mathbf{v}(\mathbf{r})$
the inequality
\begin{equation}
\label{eq4}
\int\D^3r \int \D^3r'\,
\mathbf{v}^{\ast}(\mathbf{r})\cdot
\tensor{\sigma}(\mathbf{r},\mathbf{r}',\omega)
\cdot\mathbf{v}(\mathbf{r'})>0,
\end{equation}
which is characteristic of absorbing media,
must hold.

Inserting Eq.(\ref{eq8-1}) in Eq.~(\ref{Max5}), we find that 
$\fon{E}(\mathbf{r},\omega)$ satisfies
the integro-differential equation
\begin{equation}
\label{eq8-2}
\Nabla\times\Nabla\times\fon{E}(\mathbf{r},\omega)
-\frac{\omega^2}{c^2}\,\fon{E}(\mathbf{r},\omega)
-i\mu_{0}\omega
\int\D^3r'\,\tensor{Q}(\mathbf{r,r'},\omega)
\cdot\fon{E}(\mathbf{r'},\omega)
=i\mu_{0}\omega \fon{j}_{\mathrm N}(\mathbf{r},\omega),
\end{equation}
%
which
can be regarded
as being a Langevin equation.
Note that $\fon{j}_{\mathrm N}(\mathbf{r},\omega)$,
which is a classical quantity yet,
has vanishing mean value 
(as a stochastic variable) 
and is related to the conductivity tensor through
a fluctuation--dissipation relation.
For a detailed discussion of Langevin equations,
we refer the reader to, e.\,g.,
Refs.~\cite{KuboNonEqStat,GardinerHandbook,GardinerQuantumNoise3rd}.
%
The solution to Eq.~(\ref{eq8-2}) can be
given in the form
\begin{equation}
\label{NL7}
\fon{E}(\mathbf{r},\omega)=i\mu_{0}\omega\int \D^3r'\,
\tensor{G}(\mathbf{r,r'},\omega)
\cdot\fon{j}_{\mathrm N}(\mathbf{r'},\omega),
\end{equation}
where $\tensor{G}(\mathbf{r},\mathbf{r}',\omega)$ is
the retarded Green tensor
in the frequency domain. It satisfies Eq.~(\ref{eq8-2})
with the (tensorial) $\delta$-function source,
\begin{equation}
\label{eq8-3}
\Nabla\times\Nabla\times
\tensor{G}(\mathbf{r,s},\omega)
-\frac{\omega^2}{c^2}\,\tensor{G}(\mathbf{r,s},\omega)
-i\mu_{0}\omega\int
\D^3r'\,\tensor{Q}(\mathbf{r,r'},\omega)
\cdot
\tensor{G}(\mathbf{r',s},\omega)
=
\tensor{I}\delta(\mathbf{r-s})
\end{equation}
[$\tensor{I}$, unit tensor],
together with the boundary condition
at infinity, and has all the attributes
of a Fourier transformed
response function
just as 
$\tensor{Q}(\mathbf{r},\mathbf{r}',\omega)$
has them. 
In particular, it is analytic in
the upper complex $\omega$ half-plane and the
Schwarz reflection principle
$\tensor{G}{^{\ast}}(\mathbf{r},\mathbf{r}',\omega)$
$\!=\tensor{G}(\mathbf{r},\mathbf{r}',-\omega^{\ast})$ is valid.
Since
$\tensor{Q}(\mathbf{r},\mathbf{r}',\omega)$ is reciprocal, so is
$\tensor{G}(\mathbf{r},\mathbf{r}',\omega)$,
$\tensor{G}(\mathbf{r},\mathbf{r}',\omega)\!=\!
\tensor{G}{^\mathsf{T}}(\mathbf{r}',\mathbf{r},\omega)$,
and, for real $\omega$, the generalized integral relation
\begin{equation}
\label{eq8-4}
\mu_{0}\omega
\int \D^3s\int \D^3s' \,\tensor{G}(\mathbf{r,s},\omega)
\cdot
\tensor{\sigma}(\mathbf{s,s'},\omega)
\cdot\tensor{G}{^{\ast}}(\mathbf{s',r'},\omega)
= \Im\tensor{G}(\mathbf{r,r'},\omega)
\end{equation}
holds (see Ref.~\cite{RaabeC052007}).
For absorbing media, 
Eq.~(\ref{NL7}) is the unique solution to 
Eq. (\ref{eq8-2}), i.\,e.,
it is not to be supplemented with any solutions of
the source-free version of Eq.~(\ref{eq8-2}).
Now, one can go over to the quantum-theoretical formulation
of the theory \cite{RaabeC052007}. 
Writing the operators of the
electric and magnetic induction fields
$\hat{\mathbf{E}}(\mathbf{r})$
and
$\hat{\mathbf{B}}(\mathbf{r})$, respectively,
in a picture-independent manner
as
\begin{gather}
\label{2.4b}
\hat{\mathbf{E}}(\mathbf{r})
= \int_0^\infty \D\omega\, \fo{E}(\mathbf{r},\omega) + \mathrm{H.c.},
\\
\label{2.4b-1}
\hat{\mathbf{B}}(\mathbf{r})
= \int_0^\infty \D\omega\, \fo{B}(\mathbf{r},\omega) + \mathrm{H.c.},
\end{gather}
where the respective positive-frequency parts
$\fo{E}(\mathbf{r},\omega)$ and
$\fo{B}(\mathbf{r},\omega)$
$\!=(i\omega)^{-1}$
$\!\Nabla\!\times\!\fo{E}(\mathbf{r},\omega)$
are expressed in terms of the
henceforth operator-valued
noise current density as
[cf.~Eq.~(\ref{NL7})]
\begin{gather}
\label{2.2}
\fo{E}(\mathbf{r},\omega)=i\mu_{0}\omega\int
\D^3r'\,\tensor{G}(\mathbf{r,r'},\omega)
\cdot
\fo{j}_{\mathrm N}(\mathbf{r}',\omega),
\\
\label{2.2a}
\fo{B}(\mathbf{r,\omega})=\mu_{0}
\Nabla\times \int \D^3r'\,\tensor{G}(\mathbf{r,r'},\omega)
\cdot
\fo{j}_{\mathrm N}(\mathbf{r}',\omega),
\end{gather}
one can prove (Appendix~\ref{AppComm}) that
the well known 
fundamental commutation relation
\begin{equation}
\label{eq14}
[\hat{\mathbf{E}}(\mathbf{r}),
\hat{\mathbf{B}}(\mathbf{r'})]
= i\hbar \Nabla\times\tensor{I}\delta(\mathbf{r-r'}) /
\varepsilon_{0}
\end{equation}
is satisfied if one lets
\begin{equation}
\label{eq8-6}
\bigl[\fo{j}_{\mathrm N}(\mathbf{r},\omega),
\fo{j}{^{\dagger}_{\mathrm N}}(\mathbf{r'},\omega')\bigr]
=
\frac{\hbar\omega}{\pi}
\,\tensor{\sigma}(\mathbf{r},\mathbf{r}',\omega)
\,\delta(\omega-\omega')
.
\end{equation}
To ensure that the temporal
evolution of the electromagnetic field is
in accordance with Maxwell's equations, the operators
$\fo{j}_{\mathrm N}(\mathbf{r},\omega)$ and
$\fo{j}{^{\dagger}_{\mathrm N}}(\mathbf{r},\omega)$ 
have to evolve, in the Heisenberg-picture, 
like $\sim e^{-i\omega t}$ and $\sim e^{i\omega t}$, respectively.
Hence, a Hamiltonian $\hat{H}$ 
as a functional of
$\fo{j}_{\mathrm N}(\mathbf{r},\omega)$ and
$\fo{j}{^{\dagger}_{\mathrm N}}(\mathbf{r},\omega)$ 
needs to be introduced such that
\begin{equation}
\label{NL30}
[\fo{j}_{\mathrm N}(\mathbf{r},\omega),\hat{H}]=\hbar\omega\,
\fo{j}_{\mathrm N}(\mathbf{r},\omega),
\end{equation}
which constrains the Hamiltonian 
to take the form
\begin{equation}
\label{NL28-0}
\hat H
=
\pi\!
\int_{0}^{\infty}
\D\omega\!
\int \D^3r\!
\int \D^3r'\,
\fo{j}_{\mathrm N}^{\dagger}(\mathbf{r},\omega)
\cdot
\tensor{\rho}(\mathbf{r,r'},\omega)
\cdot
\fo{j}_{\mathrm N}(\mathbf{r'},\omega),
\end{equation}
%
apart from an irrelevant $c$-number contribution.
From Eq.~(\ref{NL30}) together with
Eqs.~(\ref{eq8-6}) and (\ref{NL28-0}) one can see
that $\tensor{\rho}(\mathbf{r,r'},\omega)$ has to be chosen
to be the kernel of the integral operator that is the inverse
of the integral operator associated with
$\tensor{\sigma}(\mathbf{r,r'},\omega)$,
\begin{equation}
\label{NL-rho}
\int\D^3s\,
\tensor{\rho}(\mathbf{r,s},\omega)
\cdot
\tensor{\sigma}(\mathbf{s,r'},\omega)
=
\int\D^3s\,\tensor{\sigma}(\mathbf{r,s},\omega)
\cdot
\tensor{\rho}(\mathbf{s,r'},\omega)
=\tensor{I}\delta(\mathbf{r-r'}).
\end{equation}
Note that, by means of the correspondence
$i (\varepsilon_{0}\omega)^{-1}$
$\!\tensor{Q}(\mathbf{r},\mathbf{r}',\omega)$
$\!\leftrightarrow$
$\!\tensor{\chi}(\mathbf{r},\mathbf{r}',\omega)$,
where $\tensor{\chi}(\mathbf{r},\mathbf{r}',\omega)$ is
the (nonlocal) dielectric susceptibility tensor,
the basic commutation relation (\ref{eq8-6})
becomes equivalent to the commutation relation
derived from a microscopic, linear two-band model of
dielectric material \cite{DiStefanoO082001},
which has been used to study the quantized
electromagnetic field in spatially dispersive dielectrics
\cite{SavastaS032002,SavastaS082002}.
Provided the observables of interest---including
the Hamiltonian---can be viewed as functionals of
$\fo{j}_\mathrm N(\mathbf{r},\omega)$
[rather than functionals of the
individual current contributions in
a possible decomposition of
$\fo{j}_\mathrm N(\mathbf{r},\omega)$,
see Ref.~\cite{RaabeC052007}],
Eqs.~(\ref{eq8-6}) and (\ref{NL28-0})
can be regarded,
in view of the fluctuation-dissipation
theorem(s) (see, e.\,g., Ref.~\cite{KuboNonEqStat}),
as being unique, and hence, as invariable fundament of the theory.
If, within some frequency interval,
$\tensor{\sigma}(\mathbf{r,r'},\omega)$ does not
correspond to a positive definite integral operator
so that the integral in the inequality~(\ref{eq4})
becomes negative for suitably chosen
(quadratically integrable)
functions $\mathbf{v}(\mathbf{r})$,
linear amplification
is possible in this frequency interval.
It is not difficult to prove that
the equations given above 
can be maintained even if the positivity
condition~(\ref{eq4}) is abandoned,
provided that nevertheless
(i)~the Green tensor retains 
its analytic properties and (ii)~%
$\tensor{\rho}(\mathbf{r,r'},\omega)$ continues to
exist (for a detailed discussion of these conditions,
see Sec.~\ref{sec1-6}).
However, the diagonal form
\begin{equation}
\label{NL28}
\hat H
=
\int_{0}^{\infty}
\D\omega\, \hbar\omega
\!\int \D^3r\,
\hat{\mathbf{f}}^{\dagger}(\mathbf{r},\omega)
\cdot
\hat{\mathbf{f}}(\mathbf{r},\omega)
\end{equation}
in which the Hamiltonian (\ref{NL28-0}) can be brought in the
case of absorbing media by means of a linear (and invertible)
transformation of the variables,
\begin{equation}
\label{eq8-5}
\fo{j}_{\mathrm N}(\mathbf{r},\omega)
=
\left(\frac{\hbar\omega}{\pi}\right)^{\frac{1}{2}}
\int \D^3r'\, \tensor{K}(\mathbf{r},\mathbf{r}',\omega)
\cdot
\hat{\mathbf{f}}(\mathbf{r}',\omega),
\end{equation}
cannot be maintained if the condition~(\ref{eq4})
is abandoned.
In Eqs.~(\ref{NL28}) and (\ref{eq8-5}),
$\hat{\mathbf{f}}(\mathbf{r},\omega)$ is 
a bosonic field,
\begin{equation}
\label{eq8-7a}
\bigl[\hat{\mathbf{f}}(\mathbf{r},\omega),\hat{\mathbf{f}}^{\dagger}
(\mathbf{r'},\omega')\bigr]
=
\tensor{I}\delta(\mathbf{r}-\mathbf{r}')
\delta(\omega-\omega')
,
\end{equation}
and the integral kernel
$\tensor{K}(\mathbf{r},\mathbf{r}',\omega)$
has to obey, for real $\omega$,
the integral equation
\begin{equation}
\label{eq8-7}
\int \D^3s\,
\tensor{K}(\mathbf{r,s},\omega) \cdot
\tensor{K}{^{+}}(\mathbf{r',s},\omega)
=
\tensor{\sigma}(\mathbf{r,r'},\omega)
\end{equation}
for Eq.~(\ref{NL28}) to be an equivalent
representation of Eq.~(\ref{NL28-0}).
In the case of absorbing media, one may evidently
construct a possible integral kernel
$\tensor{K}(\mathbf{r},\mathbf{r}',\omega)$ in the form of
\begin{equation}
\label{NL21}
\tensor{K}(\mathbf{r},\mathbf{r}',\omega)
=
\int \D\alpha\,
\sigma^{\frac{1}{2}}(\alpha,\omega)\,
\mathbf{F}(\alpha,\mathbf{r},\omega)
\ot
\mathbf{F}^{\ast}(\alpha,\mathbf{r}',\omega)
\end{equation}
[$\sigma^{1/2}(\alpha,\omega)$ $\!>$ $\!0$],
where the complete and ($\delta$-)orthonormal 
functions $\mathbf{F}(\alpha,\mathbf{r},\omega)$,
\begin{equation}
\label{NL18}
\int \D\alpha\,
\mathbf{F}(\alpha,\mathbf{r},\omega)
\ot
\mathbf{F}^{\ast}(\alpha,\mathbf{r}',\omega)
=
\tensor{I}\delta(\mathbf{r}-\mathbf{r}'),
\end{equation}
\begin{equation}
\label{NL19}
\int \D^3r\,\mathbf{F}^{\ast}
(\alpha,\mathbf{r},\omega)\cdot
\mathbf{F}(\alpha',\mathbf{r},\omega)
=
\delta(\alpha-\alpha'),
\end{equation}
are defined by the eigenvalue problem
\begin{equation}
\label{NL16}
\int \D^3r'\,
\tensor{\sigma}(\mathbf{r},\mathbf{r}',\omega)
\cdot
\mathbf{F}(\alpha,\mathbf{r}',\omega)
=
\sigma(\alpha,\omega)
\mathbf{F}(\alpha,\mathbf{r},\omega).
\end{equation}
Here, the real $\omega$ plays the role of a parameter
and $\alpha$ stands for the set
of discrete and/or continuous 
quantities
needed to label the eigenfunctions. (An $\alpha$-integration 
therefore symbolizes
multiple summations and/or integrations over all those 
quantities.%
)
Note that the kernel (\ref{NL21}) is
not unique but is determined by Eq.~(\ref{eq8-7}) only
up to a unitary transformation,
which corresponds to the possibility of redefining
the dynamical variables
$\hat{\mathbf{f}}(\mathbf{r},\omega)$; for details, see
Ref.~\cite{RaabeC052007}.
Equation~(\ref{eq8-7}) is inconsistent
if $\tensor{\sigma}(\mathbf{r},\mathbf{r}',\omega)$ is not
the kernel of a positive definite integral operator so that in this case
no valid kernel $\tensor{K}(\mathbf{r},\mathbf{r}',\omega)$
exists at all.
Hence Eq.~(\ref{NL28}) 
would fail, if $\hat{\mathbf{f}}(\mathbf{r},\omega)$
were regarded as being a bosonic field satisfying the
commutation relation (\ref{eq8-7a}).


\section{Extension to linearly amplifying media}
\label{sec1-5}
In order to include in the quantization scheme linearly
amplifying media, we first note that, although
Eq.~(\ref{eq8-7}) cannot be satisfied
anymore so that Eqs.~(\ref{NL28})--(\ref{eq8-7a}) do not apply
either, Eqs.~(\ref{NL18})--(\ref{NL16})
remain valid, with $\sigma(\alpha,\omega)$
not being restricted to positive values anymore.
We may therefore expand 
$\tensor{\rho}(\mathbf{r},\mathbf{r}',\omega)$ as
\begin{equation}
\label{rho_exp}
\tensor{\rho}(\mathbf{r},\mathbf{r}',\omega)
=\int \D\alpha\, \sigma^{-1}(\alpha,\omega)
\mathbf{F}(\alpha,\mathbf{r},\omega)
\ot
\mathbf{F}^{\ast}(\alpha,\mathbf{r}',\omega),
\end{equation}
which enables us to rewrite the
Hamiltonian~(\ref{NL28-0}) as
\begin{equation}
\label{NL29-gen}
\hat H
=
\int_{0}^{\infty}
\D\omega\, \hbar\omega
\!\int\D\alpha\,
\mathrm{sgn\,} \sigma(\alpha,\omega)\,
\hat{\tilde g}^{\dagger}(\alpha,\omega)
\hat{{\tilde g}}(\alpha,\omega),
\end{equation}
where
\begin{equation}
\label{NL23-gen-inv}
\hat{{\tilde g}}(\alpha,\omega)
=
\left(\frac{\hbar\omega}{\pi}\right)^{-\frac{1}{2}}
|\sigma(\alpha,\omega)|^{-\frac{1}{2}}
\int \D^3r\,
\mathbf{F}^{\ast}(\alpha,\mathbf{r},\omega)
\cdot
\fo{j}_{\mathrm N}(\mathbf{r},\omega).
\end{equation}
With the help of Eqs.~(\ref{eq8-6}), (\ref{NL19}), and (\ref{NL16}),
it is not difficult to see that
\begin{align}
\label{NL27-gen}
\bigl[
\hat{{\tilde g}}(\alpha,\omega),
\hat{{\tilde g}}^{\dagger}(\alpha',\omega')
\bigr]
=\mathrm{sgn\,} \sigma(\alpha,\omega)\,
\delta(\alpha-\alpha')\delta(\omega-\omega').
\end{align}
The $\hat{{\tilde g}}(\alpha,\omega)$ may be viewed as non-bosonic
generalizations of the natural variables considered in
Ref.~\cite{RaabeC052007}, as they coincide with
the latter in the purely absorbing
case where 
$\mathrm{sgn\,} \sigma(\alpha,\omega)$ $\!=1$.
Inversion of Eq.~(\ref{NL23-gen-inv}) by means of Eq.~(\ref{NL18})
yields
\begin{equation}
\label{NL23-gen}
\fo{j}_{\mathrm N}(\mathbf{r},\omega)
=
\left(\frac{\hbar\omega}{\pi}\right)^{\frac{1}{2}}
\!\!
\int \D\alpha\,
|\sigma(\alpha,\omega)|^{\frac{1}{2}}\,
\mathbf{F}(\alpha,\mathbf{r},\omega)
\hat{\tilde g}(\alpha,\omega).
\end{equation}
The commutation relation (\ref{NL27-gen}) shows that
$\hat{{\tilde g}}(\alpha,\omega)$
[$\hat{{\tilde g}}^{\dagger}(\alpha,\omega)$]
is a bosonic annihilation (creation) operator for 
positive eigenvalues $\sigma(\alpha,\omega)$ 
whereas for negative ones, $\hat{{\tilde g}}(\alpha,\omega)$
[$\hat{{\tilde g}}^{\dagger}(\alpha,\omega)$]
is a creation (annihilation) operator.
It makes therefore sense to rename 
the operators according to this behavior.
Thus, denoting in each of the two cases
$\mathrm{sgn\,} \sigma(\alpha,\omega)=$ $\!\pm 1$
the respective annihilation operator by
$\hat{{b}}(\alpha,\omega)$ and the respective
creation operator by 
$\hat{{b}}^{\dagger}(\alpha,\omega)$, 
we can rewrite Eqs.~(\ref{NL29-gen}), (\ref{NL27-gen}), and
(\ref{NL23-gen}) as
\begin{equation}
\label{NL29-gen-2}
\hat H
=
\int_{0}^{\infty}
\D\omega\,\hbar\omega
\!\int\limits_{(+)}\D\alpha\,
\hat{b}^{\dagger}(\alpha,\omega)
\hat{{b}}(\alpha,\omega)
-
\int_{0}^{\infty}
\D\omega\,\hbar\omega
\!\int\limits_{(-)}\D\alpha\,
\hat{b}(\alpha,\omega)
\hat{{b}}^{\dagger}(\alpha,\omega),
\end{equation}
\begin{align}
\label{NL27-gen-2}
\bigl[
\hat{{b}}(\alpha,\omega),
\hat{{b}}^{\dagger}(\alpha',\omega')
\bigr]
=\delta(\alpha-\alpha')\delta(\omega-\omega'),
\end{align}
and
\begin{equation}
\label{NL23-gen-2}
\fo{j}_{\mathrm N}(\mathbf{r},\omega)
=
\left(\frac{\hbar\omega}{\pi}\right)^{\frac{1}{2}}
\!\!
\biggl\{
\int\limits_{(+)} \D\alpha\,
\sigma^{\frac{1}{2}}(\alpha,\omega)\,
\mathbf{F}(\alpha,\mathbf{r},\omega)
\hat{b}(\alpha,\omega)
+
\int\limits_{(-)} \D\alpha\,
[-\sigma(\alpha,\omega)]^{\frac{1}{2}}\,
\mathbf{F}(\alpha,\mathbf{r},\omega)
\hat{b}^{\dagger}(\alpha,\omega)
\biggr\}
,
\end{equation}
respectively, where the notation $\int_{(\pm)} \D\alpha\cdots$ means
that the integration extends
over those values for which
$\mathrm{sgn\,} \sigma(\alpha,\omega)=$ $\!\pm 1$.
Note that the
ranges of integration depend
on the chosen frequency in general.
It may be convenient to 
change to normal order the second term in Eq.~(\ref{NL29-gen-2}),
$\hat{b}(\alpha,\omega)\hat{{b}}^{\dagger}(\alpha,\omega)$
$\!\mapsto
\,:\hat{b}(\alpha,\omega)\hat{{b}}^{\dagger}(\alpha,\omega):$
$\!=\hat{{b}}^{\dagger}(\alpha,\omega)\hat{b}(\alpha,\omega)$,
i.\,e., to replace
the Hamiltonian in Eq.~(\ref{NL29-gen-2})
with
\begin{equation}
\label{NL29-gen-3}
\hat{H}
=
\int_{0}^{\infty}
\D\omega\, \hbar\omega
\!\int\D\alpha\,
\mathrm{sgn\,} \sigma(\alpha,\omega)\,
\hat{b}^{\dagger}(\alpha,\omega)
\hat{{b}}(\alpha,\omega),
\end{equation}
which differs from the Hamiltonian in Eq.~(\ref{NL29-gen-2})
by an (infinite but) irrelevant
$c$-number.
Since the state space of the system is to be constructed
by means of the bosonic variables $\hat{b}(\alpha,\omega)$ and
$\hat{b}^{\dagger}(\alpha,\omega)$, the
use of
Eq.~(\ref{NL29-gen-3})
in place of
Eq.~(\ref{NL29-gen-2})
is equivalent to a redefinition
(renormalization) of the zero of energy
by the condition of
absence of
quanta (vacuum state).
It should be stressed that
the temporal evolution of the variables
%
$\hat{b}(\alpha,\omega)$ and
$\hat{b}^{\dagger}(\alpha,\omega)$
which follows from
Eq.~(\ref{NL29-gen-3}) [together with Eq.~(\ref{NL27-gen-2})]
is sensitive to the sign of
$\sigma(\alpha,\omega)$ in just such a way that
Eq.~(\ref{NL23-gen-2}) always represents
the positive-frequency part of the
noise current density, as required.
From Eq.~(\ref{NL29-gen-3}) it is seen
that
there is a continuum of negative energy eigenvalues
in the case of amplification, in addition
to the positive-energy continuum
associated with absorption.
Thus, the vacuum state can no longer be
said to be the ground state of the system---there is
in fact no ground state as the continuum
stretches down to $-\infty$.
This somewhat unpleasant feature
is due to the fact that the pump mechanism which prepares 
the medium to act, in some frequency interval,
as an amplifier is not dynamically included in the theory.
Clearly, the approximation 
to treat the effect of 
pumping
within the framework of linear
response theory 
breaks down when states 
with large negative energies
are significantly involved.
By means of the transformation
\begin{align}
\label{NL25a-gen}
\,\hat{\tilde{\!\mathbf{f}}}(\mathbf{r},\omega)
&=
\int \D\alpha\,
\mathbf{F}(\alpha,\mathbf{r},\omega)
\hat{{\tilde g}}(\alpha,\omega)
\nonumber\\
&=
\int\limits_{(+)} \D\alpha\,
\mathbf{F}(\alpha,\mathbf{r},\omega)
\hat{{b}}(\alpha,\omega)
+\int\limits_{(-)} \D\alpha\,
\mathbf{F}(\alpha,\mathbf{r},\omega)
\hat{{b}}^{\dagger}(\alpha,\omega),
\end{align}
vectorial field variables
$\,\hat{\tilde{\!\mathbf{f}}}(\mathbf{r},\omega)$
can be introduced, which can be viewed as
generalizations of the variables
$\hat{\mathbf{f}}(\mathbf{r},\omega)$ 
introduced in Eq.~(\ref{eq8-5}) for the
case of purely absorbing media.
Employing Eq.~(\ref{NL19}) to invert
(the first equation in)
Eq.~(\ref{NL25a-gen}) and inserting the result in
Eq.~(\ref{NL23-gen}),
we obtain the generalization of Eq.~(\ref{eq8-5})
[$\hat{\mathbf{f}}(\mathbf{r},\omega)$
$\mapsto\,\hat{\tilde{\!\mathbf{f}}}(\mathbf{r},\omega)$],
from which we can read off,
by comparison with Eq.~(\ref{eq8-5}),
the generalization of the
kernel $\tensor{K}(\mathbf{r,r'},\omega)$,
in the form of its eigenfunction expansion.
Not surprisingly, it is
given by Eq.~(\ref{NL21}) with
$\sigma(\alpha,\omega)$ being replaced
by $|\sigma(\alpha,\omega)|$.
(It is of course possible
to consider equivalent
kernels just as in the case of purely
absorbing media.)
Unfortunately, the variables
$\,\hat{\tilde{\!\mathbf{f}}}(\mathbf{r},\omega)$
%
%
do not diagonalize the Hamiltonian in general,
and are thus less useful than in the case of 
purely absorbing media.
From Eqs.~(\ref{NL27-gen}) and (\ref{NL25a-gen}), it follows that
\begin{equation}
\label{NL27-gen-3}
[\,\hat{\tilde{\!\mathbf{f}}}(\mathbf{r},\omega),
\,\hat{\tilde{\!\mathbf{f}}}^{\dagger}
(\mathbf{r'},\omega')]
=
\delta(\omega-\omega')
\int \D\alpha\, \mathrm{sgn\,} \sigma(\alpha,\omega)\,
\mathbf{F}(\alpha,\mathbf{r},\omega)
\ot
\mathbf{F}^{\ast}(\alpha,\mathbf{r}',\omega).
\end{equation}
The integral appearing
on the right-hand side 
is the kernel of a parity-type operator, 
which reduces to the unit operator in the case of 
purely absorbing media, as it should be.
A noteworthy simplification
occurs if the medium response is strictly local---%
an assumption that is typically
made in the study of amplifying media.
In this case,
the $\,\hat{\tilde{\!\mathbf{f}}}(\mathbf{r},\omega)$
are related to the
$\hat{{\tilde g}}(\alpha,\omega)$ [$\alpha\mapsto(i,\mathbf{r})$]
in a very simple way just as in
the case of purely absorbing media that respond locally
[since the eigenfunctions
$\mathbf{F}(\alpha,\mathbf{r},\omega)$ 
are spatially localized in this case, 
see Ref.~\cite{RaabeC052007}].
Focusing, for simplicity, on isotropic media, we may then
rewrite Eqs.~(\ref{NL23-gen}) and (\ref{NL29-gen}),
respectively, as
\begin{equation}
\label{NL23-gen-3}
\fo{j}_{\mathrm N}(\mathbf{r},\omega)
=
\left(\frac{\hbar\omega}{\pi}\right)^{\frac{1}{2}}
|\sigma(\mathbf{r},\omega)|^{\frac{1}{2}}\,
\,\hat{\tilde{\!\mathbf{f}}}(\mathbf{r},\omega),
\end{equation}
and
\begin{equation}
\label{NL29-gen-4}
\hat H
=
\int_{0}^{\infty}
\D\omega\, \hbar\omega
\!\int\D^3r\,
\mathrm{sgn\,} \sigma(\mathbf{r},\omega)
\,\hat{\tilde{\!\mathbf{f}}}^{\dagger}(\mathbf{r},\omega)
\cdot
\hat{{\tilde{\!\mathbf{f}}}}(\mathbf{r},\omega),
\end{equation}
and
(\ref{NL27-gen-3}) simplifies to
\begin{align}
\label{NL27-gen-4}
\bigl[
\,\hat{{\tilde{\!\mathbf{f}}}}(\mathbf{r},\omega),
\,\hat{{\tilde{\!\mathbf{f}}}}^{\dagger}(\mathbf{r}',\omega')
\bigr]
=\mathrm{sgn\,} \sigma(\mathbf{r},\omega)\,
\tensor{I}
\delta(\mathbf{r}-\mathbf{r}')\delta(\omega-\omega').
\end{align}
Now we can switch to genuine bosonic
variables by renaming 
$\,\hat{{\tilde{\!\mathbf{f}}}}(\mathbf{r},\omega)$ 
as $\hat{\mathbf{b}}(\mathbf{r},\omega)$ and
$\hat{\mathbf{b}}^{\dagger}(\mathbf{r},\omega)$
for $\mathrm{sgn\,}\sigma(\mathbf{r},\omega)$
$\!=$ $\!1$ and 
$\mathrm{sgn\,}\sigma(\mathbf{r},\omega)$ $\!=$ $\!-1$,
respectively, so that Eq.~(\ref{NL27-gen-4}) changes to
\begin{align}
\label{NL27-gen-5}
\bigl[
\hat{\mathbf{b}}(\mathbf{r},\omega),
\hat{\mathbf{b}}^{\dagger}(\mathbf{r}',\omega')
\bigr]
=
\tensor{I}
\delta(\mathbf{r}-\mathbf{r}')\delta(\omega-\omega').
\end{align}
The noise current density
in Eq.~(\ref{NL23-gen-2}) and the
Hamiltonian in Eq.~(\ref{NL29-gen-3}), respectively,
can then be expressed in terms of
$\hat{\mathbf{b}}(\mathbf{r},\omega)$ and
$\hat{\mathbf{b}}^\dagger(\mathbf{r},\omega)$ as
\begin{equation}
\label{NL23-gen-4}
\fo{j}_{\mathrm N}(\mathbf{r},\omega)
=
\left(\frac{\hbar\omega}{\pi}\right)^{\frac{1}{2}}
\left\{
\theta[\sigma(\mathbf{r},\omega)]
\sigma^{\frac{1}{2}}(\mathbf{r},\omega)\,
\hat{\mathbf{b}}(\mathbf{r},\omega)
+
\theta[-\sigma(\mathbf{r},\omega)]
[-\sigma(\mathbf{r},\omega)]^{\frac{1}{2}}\,
\hat{\mathbf{b}}^{\dagger}(\mathbf{r},\omega)
\right\}
\end{equation}
[$\theta(x)$, unit step function] and
\begin{equation}
\label{NL29-gen-5}
\hat H
=
\int_{0}^{\infty}
\D\omega\, \hbar\omega
\!\int\D^3r\,
\mathrm{sgn\,} \sigma(\mathbf{r},\omega)
\hat{\mathbf{b}}^{\dagger}(\mathbf{r},\omega)
\cdot
\hat{\mathbf{b}}(\mathbf{r},\omega).
\end{equation}
Taking into account that
$\sigma(\mathbf{r},\omega)$ 
can be related to the
imaginary part of the
dielectric permittivity according to
$\sigma(\mathbf{r},\omega)$
$\!=\varepsilon_{0}\omega$
$\!\Im\varepsilon(\mathbf{r},\omega)$,
we find that Eq.~(\ref{NL23-gen-4}) is nothing but
the equation suggested in Ref.~\cite{ScheelS071998}
for the case of isotropic, locally and linearly responding
dielectrics that also allow for amplification.
%
%
Note that, as already mentioned for the more
general case of Eq.~(\ref{NL23-gen-2}),
both the term associated with $\hat{\mathbf{b}}(\mathbf{r},\omega)$
and the term associated with $\hat{\mathbf{b}}^\dagger(\mathbf{r},\omega)$
in Eq.~(\ref{NL23-gen-4}) give rise to
positive-frequency parts of the
noise current density.
As an illustration, let us consider the
correlation functions $\EW{\hat{\mathbf{E}}(\mathbf{r})\ot
\hat{\mathbf{E}}(\mathbf{r'})}$ and
$\EW{\hat{\mathbf{B}}(\mathbf{r})\ot
\hat{\mathbf{B}}(\mathbf{r'})}$
for the case where the system is in the vacuum state
$\ket{0}$
[$\hat{{b}}(\alpha,\omega)\ket{0}$ $\!=0$].
As well known, they play an important role in the calculation of the
Casimir force (see, e.\,g., Refs.~\cite{LifshitzEM001955,RaabeC012005}).
Using Eqs.~(\ref{2.4b})--(\ref{2.2a}) together with
Eq.~(\ref{NL23-gen-2}), taking into account Eq.~(\ref{NL27-gen-2}), 
and recalling the reciprocity of the Green tensor,
one can show by straightforward calculation that
\begin{equation}
\label{Eq8}
\EW{
\hat{\mathbf{E}}(\mathbf{r})\ot
\hat{\mathbf{E}}(\mathbf{r'})
}
=
\frac{\hbar\mu_{0}^{2}}{\pi}
\int_{0}^{\infty}\!\! \D\omega\,\omega^3
\int\D^3s
\int\D^3s'\,
\tensor{G}(\mathbf{r,s},\omega)\cdot
\tensor{\sigma}_{\mathrm{av}}
(\mathbf{s,s'},\omega)
\cdot
\tensor{G}^{\ast}(\mathbf{s',r'},\omega)
\end{equation}
and
\begin{multline}
\label{Eq9}
\bigl\langle\hat{\mathbf{B}}(\mathbf{r})\ot
\hat{\mathbf{B}}(\mathbf{r'})\bigr\rangle
\\
=
-
\frac{\hbar\mu_{0}^{2}}{\pi}
\int_{0}^{\infty}\!\! \D\omega\,\omega
\Nabla\times
\int\D^3s
\int\D^3s'\,
\tensor{G}(\mathbf{r,s},\omega)\cdot
\tensor{\sigma}_{\mathrm{av}}
(\mathbf{s,s'},\omega)
\cdot
\tensor{G}^{\ast}(\mathbf{s',r'},\omega)
\times\Lnabla{'}
,
\end{multline}
where the kernel 
$\tensor{\sigma}_{\mathrm{av}}(\mathbf{r,r'},\omega)$ reads
\begin{equation}
\label{sigmaabs}
\tensor{\sigma}_{\mathrm{av}}(\mathbf{r,r'},\omega)=
\int \D\alpha\,
|\sigma(\alpha,\omega)|
\mathbf{F}(\alpha,\mathbf{r},\omega)
\ot
\mathbf{F}^{\ast}(\alpha,\mathbf{r}',\omega).
\end{equation}
In the case of purely absorbing media, 
$\tensor{\sigma}_{\mathrm{av}}(\mathbf{r,r'},\omega)$ 
is nothing but $\tensor{\sigma}(\mathbf{r,r'},\omega)$, and by application of
the integral relation~(\ref{eq8-4}), Eqs.~(\ref{Eq8}) and (\ref{Eq9}) 
reduce to the standard
(zero-temperature) fluctuation--dissipation relations.
In contrast, if amplification is allowed for,
Eqs.~(\ref{Eq8}) and (\ref{Eq9}) differ from the standard 
formulas by terms containing the
non-vanishing difference between 
$\tensor{\sigma}_{\mathrm{av}}(\mathbf{r,r'},\omega)$ 
and 
$\tensor{\sigma}(\mathbf{r,r'},\omega)$.
In a recent study of the (zero-temperature)
Casimir force in the presence of linearly amplifying (left-handed) material
\cite{LeonhardtU082007}, the necessity of
this correction has not been taken into account. In fact,
in Ref.~\cite{LeonhardtU082007}, a 
(locally responding, isotropic)
amplifying
magnetodielectric
has been introduced,
by simply using the formulas for the correlation functions that 
are valid for absorbing media
and replacing therein the positive imaginary 
parts of the susceptibilities by negative ones. 
As we have just shown, this is wrong.

\section{Range of validity}
\label{sec1-6}
It remains to specify in more detail
the conditions that must be satisfied 
to apply the quantization scheme.
Let us first consider the
question as to whether the Green tensor
$\tensor{G}(\mathbf{r,r'},\omega)$
remains analytic
in the upper complex $\omega$ half-plane
if linear amplification is allowed for.
In the case of absorbing media, it is well-known that
solutions to the source-free macroscopic Maxwell
equations
[i.\,e., solutions to the homogeneous version of Eq.~(\ref{eq8-2})]
for real frequency 
$\omega$ can be ruled out
because of their
divergent spatial behavior.
The same is obviously `even more' true for 
frequencies in the upper 
complex $\omega$ half-plane.
Since the existence of
permissible
solutions to the source-free equations
is known to manifest itself
mathematically in the form of
singularities of the Green tensor, only the lower 
complex $\omega$ half-plane is a possible
location of such singularities
in the case of absorbing media.
In contrast, for linearly amplifying media
it may happen that permissible solutions
to the source-free macroscopic Maxwell
equations exist even if $\omega$ is chosen
in the upper complex $\omega$ half-plane.
In this case, singularities of the Green tensor
in the upper complex $\omega$ half-plane would 
exist, which would invalidate 
the proof of the commutation relation
in Eq.~(\ref{eq14}).

It is not difficult to imagine
that such `unwanted' solutions to the source-free
macroscopic Maxwell equations could arise, e.\,g.,
if waves were allowed to propagate
through an infinitely extended 
region of amplification.
Since such regions do not exist in practice,
their necessary exclusion from
consideration is of no practical relevance.
More importantly, the same
problem of `infinite amplification length'
can also occur in the case of a region of amplification of finite
extension, 
if waves can pass through the region repeatedly (due to
multiple reflections) 
with a net gain per round-trip.
Therefore,
setups where an amplifying medium is 
part of an arrangement of bodies that
act as a high-$Q$ resonator
must possibly be excluded
from consideration.
In such cases, the Green tensor
fails to be a causal function in the sense of linear
response theory.
Of course, what breaks down is not really the principle of causality but
the very concept of linear amplification---%
as fields with higher and higher
energy develop, 
the non-linear dynamics can no longer be
disregarded.
In fact, there is no need for
extra criteria to exclude from consideration
setups that would mathematically support
`unwanted' solutions---%
such setups are already excluded implicitly
by the assumption that the
approximate concept of
linear amplification is applicable.
Hence, by this basic
assumption,
the Green tensor is forced to be
analytic in the upper complex $\omega$ half-plane,
%
and
all the other important
properties of the Green tensor (in particular, its high- and
low-frequency behavior and its decay behavior for
large difference of the spatial arguments)
are the same as in the case of absorbing media.
Next, let us answer the question of the existence
of the inverse of the integral operator associated
with $\tensor{\sigma}(\mathbf{r,r'},\omega)$
in the case of linear amplification, i.\,e., the
question of
the existence of the kernel $\tensor{\rho}(\mathbf{r,r'},\omega)$
according to Eq.~(\ref{NL-rho}).
Obviously,
$\tensor{\rho}(\mathbf{r,r'},\omega)$ would
fail to exist 
if
eigenvalues
of the integral operator
associated with $\tensor{\sigma}(\mathbf{r,r'},\omega)$ were
exactly equal to zero.
Since amplification is limited to
a certain frequency range,
and since frequency is a
continuous variable,
each eigenvalue $\sigma(\alpha,\omega)$ 
on the 
negative side of the eigenvalue spectrum
can be made to move 
to the positive side by tuning the frequency,
so that zero eigenvalues are possible.
However, it should be emphasized that in the case of
a continuous spectrum this problem is in
fact harmless.
It is generally true that
zero
occurring as a continuum eigenvalue
%
does not really
preclude the inversion of (Hermitian)
operators.
A familiar example is provided by the
free-particle Schr\"{o}dinger equation
considered in the whole space. There, the continuum of $\delta$-normalizable
(plane-wave) eigenfunctions
includes the (spatially constant) zero-energy eigenfunction,
but this does not imply the nonexistence
of the inverse of the free-particle Hamiltonian
but merely the unboundedness of the inverse operator.
Although there are good reasons to believe that in practice
the eigenvalues $\sigma(\alpha,\omega)$
form a
continuous spectrum,
it seems advisable to have
also a method in store to handle
discrete zero eigenvalues,
recalling that the calculation of physical quantities
we are dealing with
generally involves space and frequency 
integrations.
As long as such
a quantity
remains meaningful as a whole, it does not really matter if
$\tensor{\rho}(\mathbf{r,r'},\omega)$ is not
literally well-defined at individual frequencies.
Thus, at least on the level of final physical
results, the effect of a discrete zero eigenvalue
cannot be
significantly different
from that of a very small but
non-zero eigenvalue
so that
any method of regularizing $\tensor{\rho}(\mathbf{r,r'},\omega)$
that implements this idea can 
be expected to give the same final results.
If this is true,
the problem that $\tensor{\rho}(\mathbf{r,r'},\omega)$ might
possibly fail to be literally well-defined 
should be irrelevant in practice.
One may simply perform all
calculations for a class of amplifying media
for which there are no problems with
$\tensor{\rho}(\mathbf{r,r'},\omega)$.
Results obtained in this way, if they are 
physically understandable, 
can then be expected to hold also
without restriction.
Finally let us address the problem of the unboundedness
from below of the energy eigenvalue spectrum
in the case of linear amplification.
Because of the lack of a lower bound,
the system could evolve
into states of lower and lower energy by the
creation of
quanta in 
the
%
frequency interval where 
$\mathrm{sgn\,} \sigma(\alpha,\omega)=$ $\!-1$, in which case 
the theory would gradually become
unrealistic.
If the system described by the Hamiltonian in
Eq.~(\ref{NL29-gen-3}) is coupled to a second system
(e.\,g., an atom),
another aspect of the problem is that
the second system might (but need not)
become more and more excited,
which of course also becomes unrealistic
at some stage.
However, such catastrophes
could only occur if the theory were
used beyond its range of validity.
In fact, they would indicate nothing but
the breakdown of the concept of linear amplification.
As long as the concept of linear amplification applies,
the unboundedness from below of the energy eigenvalue spectrum
may be regarded as being a 
purely 
formal drawback rather than a real one.
Needless to say that this unboundedness prevents one
from constructing the canonical density operator---%
the system cannot thermalize.
Therefore, it should also be clear that
electromagnetic-field correlation functions and 
fluctuation--dissipation relations in the familiar
form known for absorbing media
are not applicable to
amplifying media, not even
at zero temperature, i.\,e.,
to
the vacuum state.
Nevertheless, all the correlation functions required can be
calculated straightforwardly
for any well-defined quantum state,
in particular, for the vacuum state,
as has been
demonstrated 
at the end of Sec.~\ref{sec1-5}
for the particular 
correlation functions needed for studying the
Casimir force in the presence 
of
linearly pumped
media.
%


\section{Summary and conclusion}
\label{sec1-7}
We have shown that and how 
the very general quantization scheme developed in
Ref.~\cite{RaabeC052007}
for the macroscopic electromagnetic field in
arbitrary linearly responding media in equilibrium
can be extended
in a rather natural way
to include
media which 
(in some frequency interval 
and some spatial region)
are weakly pumped so that the effect of pumping
can be approximately described
within linear response theory.
In this sense, we have described the medium by a complex
conductivity tensor $\tensor{Q}(\mathbf{r,r'},\omega)$
and allowed for negative eigenvalues
of 
the integral operator corresponding to
its real part $\tensor{\sigma}(\mathbf{r,r'},\omega)$.
Therefore, the basic condition to apply the theory is the
validity of the concept of linear amplification
for the respective problem under study.
It has turned out that
when considering amplifying media
more care and prudence is needed than in the
case of 
robust 
equilibrium media, which only give rise to absorption.
Making contact with earlier work, we have shown that
in the special case of amplifying linear media
that can be regarded as being isotropic and locally
responding, the results in
\mbox{Refs.~\cite{ScheelS071998,MatloobR031997}}
can be recovered.
To illustrate the theory,
we have calculated the electromagnetic-field
correlation functions that determine the
Casimir force in the presence of linearly amplifying media.
As expected, they are quite different from the
well-known correlation functions in the case of absorbing
media.
Since the latter are wrongly used in Ref.~\cite{LeonhardtU082007} 
to study the effect of linearly pumped
magnetodielectrics 
on the Casimir force, with special
emphasis on left-handed metamaterials, the results
presented therein are questionable.
For a deeper understanding of the macroscopic theory
considered in the present paper, it would certainly
be advantageous to have also available
a more microscopic model
of the quantized electromagnetic
field interacting with 
linear media that also allow for amplification,
especially an analog of the well-known
Huttner-Barnett-type harmonic-oscillator models
frequently used to study
the quantized field in
absorbing media
\cite{HuttnerB101992,SuttorpLG032007}.
Such a model should
include a reservoir as in the case
of absorbing media, but presumably also
a second, `inverted' reservoir capable of being
prepared in a (formal) negative-temperature state.
To our knowledge, Huttner-Barnett-type
models that aim to incorporate amplification
have unfortunately not been developed.
%

\begin{acknowledgement}
We acknowledge discussions with 
M. Fleischhauer about the Casimir force in the
presence of linearly amplifying media.
\end{acknowledgement}


\begin{appendix}
\section{Proof of Eq.~(\ref{eq14})}
\label{AppComm}
Using Eqs.~(\ref{2.4b})--(\ref{2.2a})
and recalling the reciprocity of $\tensor{G}(\mathbf{r,r'},\omega)$,
we may write
\begin{multline}
\label{EqComm1}
\lefteqn{
[\hat{\mathbf{E}}(\mathbf{r}),
\hat{\mathbf{B}}(\mathbf{r'})]
=
-i
\mu_{0}^{2}\int_{0}^{\infty}
\D\omega\,\omega \int_{0}^{\infty}\D\omega'
\int\D^3s\int\D^3s'
}
\\
\times
\Bigl\{
\tensor{G}
(\mathbf{r,s},\omega)
\cdot
\bigl[\fo{j}_{\mathrm N}(\mathbf{s},\omega),
\fo{j}{^{\dagger}_{\mathrm N}}(\mathbf{s'},\omega')\bigr]
\cdot
\tensor{G}^{\ast}(\mathbf{s',r'},\omega')
\\
+
\tensor{G}^{\ast}(\mathbf{r,s},\omega)
\cdot
\bigl[
\fo{j}{^{\dagger}_{\mathrm N}}(\mathbf{s},\omega),
\fo{j}_{\mathrm N}(\mathbf{s'},\omega')\bigr]
\cdot
\tensor{G}(\mathbf{s',r'},\omega')
\Bigr\}
\times\Lnabla{'}.
\end{multline}
Applying the commutation relation~(\ref{eq8-6})
and employing
the reality and
reciprocity of $\tensor{\sigma}(\mathbf{r,r'},\omega)$,
we may carry out one of the frequency
integrals to obtain
\begin{multline}
\label{EqComm2}
[\hat{\mathbf{E}}(\mathbf{r}),
\hat{\mathbf{B}}(\mathbf{r'})]
=
\frac{2 \hbar\mu_{0}^{2}}{i \pi}\,
\Re\!
\int_{0}^{\infty}
\D\omega\,\omega^2
%
%
\int\D^3s\int\D^3s'
\tensor{G}
(\mathbf{r,s},\omega)
\cdot
\tensor{\sigma}
(\mathbf{s,s'},\omega)
\cdot
\tensor{G}^{\ast}(\mathbf{s',r'},\omega)
\times\Lnabla{'}.
\end{multline}
By means of the integral relation~(\ref{eq8-4}), we can
now evaluate the spatial integrals. We find,
%
on recalling the relation
$\tensor{G}^{\ast}$ $\!(\mathbf{r,r'},\omega)$ $\!=$
$\!\tensor{G}$ $\!(\mathbf{r,r'},-\omega^{\ast})$,
\begin{equation}
\label{EqComm3}
[\hat{\mathbf{E}}(\mathbf{r}),
\hat{\mathbf{B}}(\mathbf{r'})]
=
%
\frac{2 \hbar\mu_{0}}{i \pi}\,
\int_{0}^{\infty}
\D\omega\,\omega\,
\Im
\tensor{G}
(\mathbf{r,r'},\omega)
\times\Lnabla{'}
%
=
%
\frac{\hbar\mu_{0}}{i \pi}\,
\left[
\int_{-\infty}^{\infty}
\D\omega\,\omega\,
\Im
\tensor{G}
(\mathbf{r,r'},\omega)
\right]
\times\Lnabla{'}
.
\end{equation}
As, within the framework of macroscopic QED,
the presence of any medium is irrelevant at sufficiently
high frequencies,
the leading asymptotic behavior of
$\tensor{G}(\mathbf{r,r'},\omega)$ for 
frequencies
in the upper $\omega$ half-plane (including the real axis)
%
with large absolute values
is given by
\begin{equation}
\label{EqComm4}
\tensor{G}(\mathbf{r,r'},\omega)
\simeq -\frac{c^2}{\omega^2}\tensor{I}\delta(\mathbf{r}-\mathbf{r'}),
\end{equation}
%
just as for the Green tensor in free space.
Together with the fact that
the most singular term
of $\tensor{G}(\mathbf{r,r'},\omega)$
at $\omega$ $\!=0$
is $-c^{2}\tensor{L}(\mathbf{r,r'})/\omega^{2}$,
where $\tensor{L}(\mathbf{r,r'})$ is a real tensor that is purely
longitudinal from both sides
[and thus does not contribute to
$\Im\tensor{G}(\mathbf{r,r'},\omega)$ for real $\omega$],
this asymptotic behavior shows that the integral
in the
square brackets
in Eq.~(\ref{EqComm3}) converges.
It may be evaluated by contour-integral techniques as
\begin{align}
\label{EqComm5}
\int_{-\infty}^{\infty}
\D\omega\,\omega\,
\Im
\tensor{G}
(\mathbf{r,r'},\omega)
&
=
\Im
\mathcal{P}
\int_{-\infty}^{\infty}
\D\omega\,\omega
\tensor{G}
(\mathbf{r,r'},\omega)
\nonumber\\
&=
\Im
\int_{\mathcal{C}}
\D\omega\,\omega
\tensor{G}
(\mathbf{r,r'},\omega)
%
=
\pi c^{2}
[\tensor{I}
\delta(\mathbf{r}-\mathbf{r'})
-\tensor{L}
(\mathbf{r,r'})].
\end{align}
Here,
the principal-value $(\mathcal{P})$
integral
has been changed to an integral over
a contour $\mathcal{C}$ that consists of an infinitely large semi-circle
in the upper half-plane (traversed clockwise), plus an infinitely
small semi-circle (traversed counter-clockwise)
that avoids the origin in the upper half-plane.
Note that the sub-leading (weaker than $\omega^{-2}$)
singular terms of
$\tensor{G}(\mathbf{r,r'},\omega)$
at $\omega$ $\!=0$
do not contribute,
irrespective of the actual nature of the 
singularity.
%
Inserting Eq.~(\ref{EqComm5}) in Eq.~(\ref{EqComm3}) and using
$\tensor{L}(\mathbf{r,r'})\times\Lnabla{'}$ $\!=0$,
we arrive at the desired Eq.~(\ref{eq14})
[$\Nabla\times\tensor{I}\delta(\mathbf{r-r'})$
$\!=-\tensor{I}\delta(\mathbf{r-r'})\times\Lnabla{'}$].
\end{appendix}


%
\end{document}